\documentclass[12pt]{article}
\usepackage{latexsym}
\usepackage{amsmath,amsfonts}
\usepackage{times}
\allowdisplaybreaks[4]

\catcode`\@=11

\newcount\hour
\newcount\minute
\newtoks\amorpm \hour=\time\divide\hour by 60\minute
=\time{\multiply\hour by 60 \global\advance\minute by-\hour}
\edef\standardtime{{\ifnum\hour<12 \global\amorpm={am}%
        \else\global\amorpm={pm}\advance\hour by-12 \fi
        \ifnum\hour=0 \hour=12 \fi
        \number\hour:\ifnum\minute<10
        0\fi\number\minute\the\amorpm}}
\edef\militarytime{\number\hour:\ifnum\minute<10 0\fi\number\minute}

\def\draftlabel#1{{\@bsphack\if@filesw {\let\thepage\relax
   \xdef\@gtempa{\write\@auxout{\string
      \newlabel{#1}{{\@currentlabel}{\thepage}}}}}\@gtempa
   \if@nobreak \ifvmode\nobreak\fi\fi\fi\@esphack}
        \gdef\@eqnlabel{#1}}
\def\@eqnlabel{}
\def\@vacuum{}
\def\marginnote#1{}
\def\draftmarginnote#1{\marginpar{\raggedright\scriptsize\tt#1}}
\def\draft{
        \pagestyle{plain}
        \overfullrule=2pt
        \oddsidemargin -.5truein
        \def\@oddhead{\sl \phantom{\today\quad\militarytime} \hfil
        \smash{\Large\sl DRAFT} \hfil \today\quad\militarytime}
        \let\@evenhead\@oddhead
        \let\label=\draftlabel
        \let\marginnote=\draftmarginnote
        \def\ps@empty{\let\@mkboth\@gobbletwo
        \def\@oddfoot{\hfil \smash{\Large\sl DRAFT} \hfil}
        \let\@evenfoot\@oddhead}
        \def\@eqnnum{(\theequation)\rlap{\kern\marginparsep\tt\@eqnlabel}%
        \global\let\@eqnlabel\@vacuum}  }

\newcommand{\rf}[1]{(\ref{#1})}
\renewcommand{\theequation}{\thesection.\arabic{equation}}
\renewcommand{\thefootnote}{\fnsymbol{footnote}}
\newcommand{\newsection}{   
\setcounter{equation}{0}\section}

\def\appendix#1{\addtocounter{section}{1}\setcounter{equation}{0}
\renewcommand{\thesection}{\Alph{section}}
\section*{Appendix \thesection\protect\indent \parbox[t]{11.15cm}{#1}}
\addcontentsline{toc}{section}{Appendix \thesection\ \ \ #1}}

\def\be{\begin{equation}}
\def\ee{\end{equation}}
\def\beq{\begin{eqnarray}}
\def\eeq{\end{eqnarray}}

\def\Boxline{{\Box \kern-0.7em  /  }\, }
\def\Dline{{D \kern-0.6em  /  }\, }
\def\parline{\,\partial\kern -0.55em /\,\,}

\def\half{{\frac{1}{2}}}

\def\DD{{\cal D}}

\def\LL{{\cal L}}

\def\ibf{{\bf i}}
\def\iibf{{\bf ii}}
\def\iiibf{{\bf iii}}

\def\psik{|\psi\rangle}
\def\psibr{\langle\psi|}

\def\xik{|\xi\rangle}
\def\Xik{|\Xi\rangle}

\def\Gammasm{{\scriptscriptstyle{\Gamma}}}
\def\Ism{{\scriptscriptstyle{I}}}
\def\IIsm{{\scriptscriptstyle{II}}}

\def\Gammasm{{\scriptscriptstyle \Gamma}}

\def\smzero{{\scriptscriptstyle (0)}}
\def\smone{{\scriptscriptstyle (1)}}

\def\smzero{{\scriptscriptstyle (0)}}
\def\smone{{\scriptscriptstyle (1)}}

\def\smponetwo{{\scriptscriptstyle [1,2]}}

\def\sm(A)dS{{\scriptscriptstyle (A)dS }}

\def\Ssm{{\scriptscriptstyle S }}

\def\gaal{\gamma\alpha}

\def\alphab{\bar\alpha}
\def\upsilonb{\bar\upsilon}
\def\gaalb{\gamma\bar\alpha}

\def\Cb{\bar{C}}
\def\eb{\bar{e}}

\def\irm{{\rm i}}

\def\st{{\rm st}}

\def\st{{\rm st}}
\def\mod{{\rm mod}}
\def\bos{{\rm bos}}

\def\smponetwo{{\scriptscriptstyle [1,2]}}

\def\mun{{\underline{m}}}

\begin{document}


\begin{flushright}
FIAN-TD-2017-06 \hspace{1.8cm}  {}~\\
arXiv: 1703.05780 V2 [hep-th] \\
\end{flushright}

\vspace{1cm}

\begin{center}

{\Large \bf Fermionic continuous spin gauge field in (A)dS space}

\vspace{2.5cm}

R.R. Metsaev%
\footnote{ E-mail: metsaev@lpi.ru
}

\vspace{1cm}

{\it Department of Theoretical Physics, P.N. Lebedev Physical
Institute, \\ Leninsky prospect 53,  Moscow 119991, Russia }

\vspace{3.5cm}

{\bf Abstract}

\end{center}

Fermionic continuous spin field propagating in (A)dS space-time is studied.
Gauge invariant Lagrangian formulation for such fermionic field is developed.
Lagrangian of the fermionic continuous spin field is constructed in terms of triple gamma-traceless tensor-spinor Dirac fields, while gauge symmetries are realized by using gamma-traceless gauge transformation parameters. It is demonstrated that partition function of fermionic continuous spin field is equal to one. Modified de Donder gauge condition that considerably simplifies analysis of equations of motion is found. Decoupling limits leading to arbitrary spin massless, partial-massless, and massive fermionic fields are studied.

\vspace{2cm}

Keywords: Fermionic field; Continuous spin; Higher-spin field.

\newpage
\renewcommand{\thefootnote}{\arabic{footnote}}
\setcounter{footnote}{0}

\newsection{\large Introduction}

In view of many interesting features of continuous spin gauge field theory this topic has attracted some interest in recent time \cite{Najafizadeh:2015uxa}-\cite{Metsaev:2016lhs}. For a list of references devoted to various aspects of continuous spin field, see Refs.\cite{Schuster:2014hca,Bekaert:2005in,Brink:2002zx}.
It seems likely that some regime of the string theory can be
related to continuous spin field theory (see, e.g., Refs.\cite{Savvidy:2003fx}). Other interesting feature of continuous spin field theory, which triggered our interest in this topic, is that the bosonic continuous spin field can be decomposed into an infinite chain of coupled scalar, vector, and totally symmetric tensor fields which consists of every field just once. We recall then that a similar infinite chain of scalar, vector and totally symmetric fields enters the theory of bosonic higher-spin gauge field in AdS space \cite{Vasiliev:1990en}.

Supersymmetry plays important role in string theory and higher-spin gauge field theory. We expect that supersymmetry will also play important role in theory of continuous spin field. For a discussion of supersymmetry, we need to study bosonic and fermionic continuous spin fields. Lagrangian formulation of bosonic continuous spin field in flat space $R^{3,1}$ was studied in Ref.\cite{Schuster:2014hca}, while Lagrangian formulation of bosonic continuous spin field in flat space $R^{d,1}$ and $(A)dS_{d+1}$ space with arbitrary $d$ was discussed in Ref.\cite{Metsaev:2016lhs}. Lagrangian formulation of fermionic continuous spin field in flat space $R^{3,1}$ was obtained in Ref.\cite{Najafizadeh:2015uxa}. So far Lagrangian formulation of fermionic continuous spin field in $(A)dS_{d+1}$ and $R^{d,1}$ with arbitrary $d$ has not been discussed in the literature.
Our major aim in this paper is to develop Lagrangian formulation of continuous spin fermionic field in flat space $R^{d,1}$ and  $(A)dS_{d+1}$ space with arbitrary $d\geq 3$.%
\footnote{ We agree with remarks of Authors in Refs.\cite{Najafizadeh:2015uxa,Schuster:2014hca} that generalization of Lagrangian for continuous spin massless field in flat space $R^{3,1}$ to the case of $R^{d,1}$ with $d>3$ is straightforward.}
As by product, using our Lagrangian formulation, we compute partition functions
of fermionic continuous spin fields in (A)dS and flat spaces and show that such partition functions are equal to 1. Considering various decoupling limits, we demonstrate how massless, partial-massless, and massive fermionic fields appear in the framework of Lagrangian formulation of fermionic continuous spin (A)dS field.

\newsection{ \large Lagrangian formulation of fermionic continuous spin field}

{\bf Field content}. To discuss gauge invariant and Lorentz covariant formulation of fermionic continuous spin field propagating in $(A)dS_{d+1}$ space we introduce the following set of Dirac complex-valued
spin-$\half$, spin-$\frac{3}{2}$, and tensor-spinor fields of the Lorentz $so(d,1)$
algebra
\be \label{man-11022017-01}
\psi^{a_1\ldots a_n\alpha}\,, \qquad\qquad n =0,1,\ldots, \infty\,,
\ee
\vspace{-0.1cm}
where flat vector indices  of the $so(d,1)$ algebra $a_1,\ldots,a_n$ run over $0,1,\ldots,d$, while $\alpha$ stands for spinor index. In what follows, the spinor indices will be implicit.  In \rf{man-11022017-01}, fields with $n=0$ and $n=1$ are the respective Dirac spin-$\half$ and spin-$\frac{3}{2}$ fields of the $so(d,1)$ algebra, while fields with $n \geq 2$ are the totally symmetric spin-$(n+\half)$ Dirac tensor-spinor fields of the $so(d,1)$ algebra. Tensor-spinor fields $\psi^{a_1\ldots a_n}$ \rf{man-11022017-01} with $n \geq 3$ are considered to be triple gamma-traceless,%
\be \label{man-11022017-02}
\gamma^a\psi^{abba_4\ldots a_n}=0\,, \hspace{2cm}
n=3,4,\ldots,\infty.
\ee
\vspace{-0.1cm}
Dirac fields \rf{man-11022017-01} subject to constraints \rf{man-11022017-02}
constitute a field content in our approach.

In order to obtain a gauge invariant description of the continuous spin fermionic field
in an easy--to--use form, we introduce bosonic creation operators $\alpha^a$, $\upsilon$. Using the $\alpha^a$, $\upsilon$, we collect all fields appearing in \rf{man-11022017-01} into a ket-vector $|\psi\rangle$,%
\be \label{man-11022017-03}
\psik = \sum_{n=0}^\infty \frac{\upsilon^n}{n!\sqrt{n!}} \alpha^{a_1} \ldots \alpha^{a_n} \psi^{a_1\ldots a_n} |0\rangle\,.
\ee
We note that triple gamma-tracelessness constraint \rf{man-11022017-02} can be represented as $\alphab^2 \gaalb \psik = 0$.

\noindent {\bf Lagrangian}. Gauge invariant action and Lagrangian of the fermionic continuous spin field we found take the form
\beq
\label{man-11022017-04} S  & = &  \int d^{d+1}x\,\LL\,,  \qquad  \irm\LL  = e \psibr   E \psik \,,
\\
\label{man-11022017-05} && E  = E_\smone + E_\smzero\,,
\\
\label{man-11022017-06}  && E_\smone = \Dline  - \alpha D \gamma\bar\alpha -
\gamma\alpha\bar\alpha D + \gamma\alpha \Dline\gamma\bar\alpha +
\frac{1}{2}\gamma\alpha\alpha D \bar\alpha^2 + \frac{1}{2}\alpha^2\gamma\bar\alpha\bar\alpha D -
\frac{1}{4}\alpha^2\Dline\bar\alpha^2\,, \qquad
\\
\label{man-11022017-07}  && E_\smzero  = (1 - \gamma\alpha\gamma\bar\alpha - \frac{1}{4}\alpha^2\bar\alpha^2) e_1^\Gammasm + (\gamma\alpha - \half\alpha^2\gamma\bar\alpha)\eb_1 + (\gamma\bar\alpha -\half \gamma\alpha\bar\alpha^2) e_1\,,
\eeq
$\langle\psi| \equiv (\psik)^\dagger\gamma^0$, where $e=\det e_\mun^a$ and $e_\mun^a$ stands for vielbein in (A)dS space. In \rf{man-11022017-06} and below, the notation  $\Dline$ is used for the  Dirac operator in (A)dS space. A definition of quantities like $\alpha D$, $\gaal$, $\alpha^2$ may be found in Appendix. Quantities $e_1^\Gammasm$, $e_1$, and $\eb_1$ entering $E_\smzero$ \rf{man-11022017-07} are defined as
\beq
\label{man-11022017-08} && e_1^\Gammasm  = \frac{2\kappa_0}{2N_\upsilon + d-1}\,, \qquad e_1 =  e_\upsilon \upsilonb\,, \qquad\qquad  \eb_1 = - \upsilon e_\upsilon\,,
\\
\label{man-11022017-09} && e_\upsilon  = \Bigl(\frac{F_\upsilon }{(N_\upsilon + 1) (2N_\upsilon+d-1)} \Bigr)^{1/2} \,, \hspace{1.4cm} N_\upsilon = \upsilon\upsilonb\,,
\\
\label{man-11022017-10} && F_\upsilon =  \kappa_0^2 -  \mu_0 \bigl(N_\upsilon + \frac{d-1}{2} \bigr)^2  - \rho \bigl(N_\upsilon + \frac{d-1}{2} \bigr)^4 \,,
\\
\label{man-11022017-10-a1} && \kappa_0 \ \ \hbox{and} \ \ \mu_0 - \hbox{ dimensionfull parameters},
\\
\label{man-11022017-10-a2} && \rho = - R^{-2} \ \hbox{ for AdS}; \quad \rho = 0 \ \hbox{ for flat}; \quad \rho = R^{-2} \ \hbox{ for dS},
\eeq
where the $R$ stands for a radius of (A)dS space. The following remarks are in order.\\
\noindent \ibf) Our Lagrangian depends on two arbitrary dimensionfull parameters $\kappa_0$, $\mu_0$ and on $\rho$ \rf{man-11022017-10-a2}.

\noindent \iibf) The one-derivative operator $E_\smone$ \rf{man-11022017-06} coincides with the one-derivative contribution to the Fang-Fronsdal operator that enters Lagrangian
of fermionic massless field in (A)dS space.

\noindent {\bf Gauge symmetries}. In order to describe gauge transformation of continuous spin field we use the following gauge transformation parameters:
\be \label{man-11022017-12}
\xi^{a_1\ldots a_n \alpha}\,,\qquad\qquad n=0,1,\ldots, \infty\,,
\ee
where $\alpha$ stands for spinor index which will be implicit in what follows.  In \rf{man-11022017-12}, gauge transformation parameters with $n=0$ and $n=1$ are the
respective spin-$\half$ and spin-$\frac{3}{2}$ Dirac fields of the Lorentz $so(d,1)$ algebra, while  gauge transformation parameters with $n\geq 2$ are totally symmetric $\gamma$-traceless spin-$(n+\half)$ Dirac fields of the Lorentz $so(d,1)$ algebra,
\be \label{man-11022017-14}
\gamma^a\xi^{aa_2\ldots a_n}=0\,, \qquad n= 1,2,\ldots, \infty\,.
\ee
In order to simplify the presentation of gauge transformation we use the oscillators $\alpha^a$, $\upsilon$ and collect all gauge transformation parameters appearing in \rf{man-11022017-12} into ket-vector $\xik$ defined by the relation
\be \label{man-11022017-15}
\xik = \sum_{n=0}^\infty \frac{\upsilon^{n+1}}{n!\sqrt{(n+1)!}} \alpha^{a_1} \ldots \alpha^{a_n} \xi^{a_1\ldots a_n} |0\rangle\,.
\ee
In terms of the ket-vector $\xik$,
$\gamma$-tracelessness constraints \rf{man-11022017-12}
are represented as $\gaalb \xik=0$.

We now find that our Lagrangian \rf{man-11022017-04}-\rf{man-11022017-07} is invariant under a gauge transformation given by
\be \label{man-11022017-16}
\delta \psik  = G\xik\,, \qquad G = \alpha D - e_1 + \gaal\frac{e_1^\Gammasm}{2N_\alpha
+ d- 1} - \alpha^2\frac{1}{2N_\alpha + d + 1}\eb_1 \,,
\ee
where the operators $e_1^\Gammasm$, $e_1$, $\eb_1$ entering derivative independent part of gauge transformation \rf{man-11022017-16} are defined in \rf{man-11022017-08}-\rf{man-11022017-10-a2}.

Representation for gauge invariant Lagrangian \rf{man-11022017-04}-\rf{man-11022017-07} and the corresponding gauge transformation \rf{man-11022017-16} is universal and valid for arbitrary theory of fermionic totally symmetric gauge (A)dS fields. Various
theories of fermionic totally symmetric gauge (A)dS fields are distinguished only by explicit form of the operators $e_1^\Gammasm$, $e_1$, $\eb_1$ entering $E_\smzero$ \rf{man-11022017-07} and gauge transformation \rf{man-11022017-16}. This is to say that, operators $E$ and $G$ for fermionic totally symmetric massless, massive, conformal, and continuous spin fields in (A)dS depend on the oscillators $\alpha^a$, $\alphab^a$, the Dirac $\gamma$-matrices, and on the derivative $D^a$  in the same way as operators $E$ \rf{man-11022017-05}-\rf{man-11022017-07} and $G$ \rf{man-11022017-16}. Namely, operators $E$ and $G$ for fermionic totally symmetric massless, massive, conformal, and continuous spin fields in (A)dS are distinguished only by the explicit form of the operators $e_1^\Gammasm$, $e_1$, and $\eb_1$. For the reader convenience we note that, for massless fields in $(A)dS_{d+1}$, the $e_1^\Gammasm$, $e_1$, and $\eb_1$ take the form
\vspace{-0.1cm}
\be \label{man-11022017-17}
e_1^\Gammasm = \sqrt{-\rho} \bigl(N_\upsilon + \frac{d-3}{2}\bigr)\,, \qquad e_1  = 0, \qquad \eb_1= 0, \quad \hbox{ for massless fields in } (A)dS_{d+1}\,.
\ee
For spin-$(s+\half)$ and mass-m massive (A)dS field, the operators $e_1^\Gammasm$, $e_1$, and $\eb_1$ can be read from expressions (2.18)-(2.21) in Ref.\cite{Metsaev:2006zy}.%
\footnote{
In this paper, we use  metric-like approach to gauge fields.
Let us also mention frame-like and BRST approaches to gauge fields (see, e.g., Refs.\cite{Alkalaev:2005kw,Buchbinder:2007vq}). It will be interesting to study fermionic continuous spin field by using frame-like and BRST approaches and establish their connection with a vector-superspace approach  in Refs.\cite{Najafizadeh:2015uxa,Schuster:2014hca,Rivelles:2014fsa}. Study of bosonic continuous spin field by using BRST method may be found in Ref.\cite{Bengtsson:2013vra}.
}
For fermionic conformal fields in flat space, the operators $e_1^\Gammasm$, $e_1$, and $\eb_1$ can be read from expressions (3.31)-(3.34) in Ref.\cite{Metsaev:2012hr}.
For fermionic conformal fields in (A)dS space, explicit expressions for the operators $e_1^\Gammasm$, $e_1$, and $\eb_1$ are still to be worked out.

\newsection{\large  (Ir)reducible classically unitary fermionic continuous spin field}

Our Lagrangian \rf{man-11022017-04} depends on the two parameters $\kappa_0$ and $\mu_0$. In this Section, we find restrictions imposed on the $\kappa_0$ and $\mu_0$ for irreducible and reducible classically unitary dynamical systems.  Let us start with our definition of classically unitary reducible and irreducible systems.

\noindent \ibf) Lagrangian \rf{man-11022017-04} is constructed out of complex-valued Dirac fields. In order for the action be hermitian the $\kappa_0$ \rf{man-11022017-08} should be real-valued, while the quantity $F_\upsilon$ \rf{man-11022017-10} should be positive for all eigenvalues $N_\upsilon=0,1,\ldots,\infty$. Introducing the notation
\beq
\label{man-19022017-14} && F_\upsilon(n)  =  \kappa_0^2 -  \mu_0 \bigl( n + \frac{d-1}{2} \bigr)^2  - \rho \bigl( n + \frac{d-1}{2} \bigr)^4 \,, \qquad F_\upsilon(n) \equiv F_\upsilon|_{N_\upsilon=n}\,,
\eeq
we note then that, depending on behaviour of the $F_\upsilon(n)$, we use the following terminology
\beq
\label{man-19022017-15}  && F_\upsilon(n)\geq 0 \ \hbox{ for all } \ n=0,1,\ldots, \infty  \hspace{1.8cm} \hbox{ classically unitary system};\qquad
\\
\label{man-19022017-16}  && F_\upsilon(n) \ne 0 \ \hbox{ for all } \ n =  0,1,\ldots, \infty,  \hspace{1.7cm} \hbox{ irreducible system};
\\
\label{man-19022017-17}  && F_\upsilon(n_r) = 0 \ \hbox{ for some } \ n_r \in 0,1,\ldots, \infty,  \hspace{1cm} \hbox{ reducible system}.
\eeq
Relation \rf{man-19022017-15} tells us that, if $F_\upsilon(n)$ \rf{man-19022017-14} is positive for all $n$, then we will refer to fields \rf{man-11022017-01} as classically unitary system. From \rf{man-19022017-16}, we learn that, if $F_\upsilon(n)$ \rf{man-19022017-14} has no roots,  then we will refer to fields \rf{man-11022017-01} as irreducible system. For this case, Lagrangian \rf{man-11022017-04} describes infinite chain of coupling fields \rf{man-11022017-01}. From \rf{man-19022017-17}, we learn  that, if $F_\upsilon(n)$ \rf{man-19022017-14} has roots, then we will refer to fields \rf{man-11022017-01} as reducible system. For the reducible system, Lagrangian \rf{man-11022017-04} and gauge transformation \rf{man-11022017-16} are factorized and describe finite and infinite decoupled chains of fields.

From now on we assume that $\kappa_0$ is real-valued. Using definitions above-given in \rf{man-19022017-15}-\rf{man-19022017-17}, we define (ir)reducible classically unitary systems as follows
\beq
\label{man-19022017-18}  && \hspace{-1.4cm} F_\upsilon(n) > 0 \ \hbox{ for all } \ n =0,1,\ldots, \infty,  \hspace{2.8cm} \hbox{ irreducible classically unitary system};\qquad
\\
&& \hspace{-1.4cm}  F_\upsilon(n_r) = 0  \ \hbox{ for some } n_r \in  0,1,\ldots, \infty\,,
\nonumber\\
\label{man-19022017-20} && \hspace{-1.4cm} F_\upsilon(n) > 0  \ \hbox{ for all } n =0,1,\ldots, \infty \hbox{ and } n \ne n_r \hspace{1cm} \hbox{ reducible classically unitary system}.
\eeq

We now consider (ir)reducible classically unitary systems for flat, AdS, and dS spaces in turn.

\noindent {\bf Flat space, $\rho=0$}. Setting $\rho=0$ in \rf{man-19022017-14}, it is easy to see that restrictions \rf{man-19022017-18},\rf{man-19022017-20} amount to the following restrictions on the parameters $\kappa_0$ and $\mu_0$:
\beq
\label{man-27092016-06} && \hspace{-1cm} \mu_0 = 0\,, \qquad \kappa_0 \ne 0\,, \hspace{2.3cm}  \hbox{ for irreducible classically unitary system},
\\
\label{man-27092016-06-a2} && \hspace{-1cm}  \mu_0 < 0\,, \qquad \kappa_0 - \hbox{ arbitrary }\,, \qquad  \hbox{ for irreducible classically unitary system},\qquad
\\
\label{man-27092016-07} && \hspace{-1cm} \mu_0 =0\,, \qquad \kappa_0 = 0\,,  \hspace{2.3cm}   \hbox{ for reducible classically unitary system}.
\eeq
Below, we demonstrate that $\mu_0$ is related to conventional mass parameter $m$ as $\mu_0=m^2$ (see \rf{man-13032017-04} for $\rho=0$). This implies that the cases $\mu_0 = 0$, $\mu_0 > 0$, and $\mu_0 < 0 $ are associated with the respective massless, massive, and tachyonic fields.
Note that, for the case of \rf{man-27092016-07}, we have $F_\upsilon(n)\equiv 0$ and this case describes a chain of conventional massless fields in flat space which consists of every spin just once. Case $\mu_0=0$ \rf{man-27092016-06} describes massless continuous spin field. For such field in $R^{3,1}$, alternative representation for gauge invariant Lagrangian was obtained first in  Ref.\cite{Najafizadeh:2015uxa}. Case $\mu_0<0$ \rf{man-27092016-06-a2} describes tachyonic continuous spin field in flat space. Lagrangian gauge invariant description for fermionic massless continuous spin field in $R^{d,1}$, $d>3$, and fermionic tachyonic continuous spin field in $R^{d,1}$, $d\geq 3$, has not been discussed in earlier literature. We expect that our continuous spin field with $\mu_0 < 0$ \rf{man-27092016-06-a2} is associated with tachyonic UIR of the Poincar\'e algebra. Tachyonic UIR of the Poincar\'e algebra are discussed, e.g., in Ref.\cite{Bekaert:2006py}.

Ignoring restrictions \rf{man-19022017-15}, \rf{man-19022017-16}, we find that restriction \rf{man-19022017-17} leads to interesting reducible system. Namely, setting $\rho=0$ in \rf{man-19022017-14}, we see that the equation $F_\upsilon(s)=0$ gives
\be \label{man-19022017-21}
\kappa_0^2 = \bigl(s+ \frac{d-1}{2}\bigr)^2 \mu_0\,.
\ee
Inserting $\rho=0$ and $\kappa_0^2$ \rf{man-19022017-21} into \rf{man-19022017-14}, we get
\be \label{man-19022017-22}
F_\upsilon(n) =   (s-n)(s + n +d-1)  \mu_0  \,.
\ee
Now decomposing ket-vectors $\psik$,  \rf{man-11022017-03} and $\xik$ \rf{man-11022017-15} as
\beq
\label{man-19022017-23} \psik & = &  |\psi^{0,s}\rangle  + |\psi^{s+1,\infty}\rangle\,,
\\
\label{man-19022017-24} && \hspace{0cm} |\psi^{M,N}\rangle \equiv  \sum_{n=M}^N \frac{\upsilon^n}{n!\sqrt{n!}} \alpha^{a_1} \ldots \alpha^{a_n} \psi^{a_1\ldots a_n} |0\rangle\,,
\\
\label{man-19022017-25}   \xik & = & |\xi^{0,s-1}\rangle  + |\xi^{s,\infty}\rangle \,,
\\
\label{man-19022017-26} &&  \hspace{0cm}  |\xi^{M,N}\rangle \equiv \sum_{n=M}^N \frac{\upsilon^{n+1}}{n!\sqrt{(n+1)!}} \alpha^{a_1} \ldots \alpha^{a_n} \xi^{a_1\ldots a_n} |0\rangle\,,
\eeq
and using $\rho=0$ and $F_\upsilon$ as in \rf{man-19022017-22},  we verify that Lagrangian \rf{man-11022017-04} and gauge transformation \rf{man-11022017-16} are factorized,
\beq
\label{man-19022017-27} && \hspace{-1.5cm} \LL = \LL^{0,s} + \LL^{s+1,\infty}\,, \qquad \irm \LL^{0,s} \equiv \langle \psi^{0,s}|E|\psi^{0,s}\rangle\,,  \qquad \irm \LL^{s+1,\infty} \equiv
\langle \psi^{s+1,\infty}|E|\psi^{s+1,\infty}\rangle\,,\qquad
\\
\label{man-19022017-28} && \hspace{3cm} \delta |\psi^{0,s}\rangle  = G|\xi^{0,s-1}\rangle\,, \qquad \ \delta |\psi^{s+1,\infty}\rangle  = G |\xi^{s,\infty}\rangle\,,
\eeq
where $E$, $G$ are given in \rf{man-11022017-05},\rf{man-11022017-16}. Thus, for values of the parameters $\mu_0$ and $\kappa_0$ given in \rf{man-19022017-21}, we see from \rf{man-19022017-27}, \rf{man-19022017-28} that our Lagrangian and gauge transformations describe two decoupling fields $|\psi^{0,s}\rangle$ and  $|\psi^{s+1,\infty}\rangle$. As $\kappa_0$ \rf{man-19022017-21} is real-valued, we have $\mu_0> 0$. Using the notation $\mu_0=m^2$, we note that the field $|\psi^{0,s}\rangle$ describes a fermionic classically unitary spin-$(s+\half)$  mass-$m$ massive field. For this case, $F_\upsilon(n) < 0$ \rf{man-19022017-22} when $n=s+1,s+2,\ldots,\infty$ and this implies that the ket-vector $|\psi^{s+1,\infty}\rangle$ describes classically non-unitary fermionic field.

\noindent {\bf (A)dS space, $\rho \ne 0$}. For (A)dS, our study of Eqs.\rf{man-19022017-18},\rf{man-19022017-20} is summarized as follows.

\noindent {\bf Statement 1}. {\it For dS, equations \rf{man-19022017-18}, \rf{man-19022017-20} do not have solutions}. This implies that continuous spin dS field is not realized neither as irreducible  classically unitary  system nor as reducible classically unitary system.%
\footnote{ Discussion of group theoretical aspects of quantum fields in dS space may be found in Refs.\cite{Joung:2006gj,Basile:2016aen}.}

\noindent {\bf Statement 2}. {For AdS, Eqs.\rf{man-19022017-18} have solutions which we classify as Type IA,IB, II, and III solutions}.
{\small\\
\noindent {\it Type IA and IB solutions for AdS:}
\beq
\label{man-12032017-01} && \hspace{-2cm} \mu_0 < \frac{1}{4} |\rho| (d-1)^2 \,, \hspace{3.5cm} \kappa_0- \hbox{arbitrary}, \hspace{4.4cm} IA;
\\
\label{man-12032017-02} && \hspace{-2cm} \frac{1}{4} |\rho| (d-1)^2 \leq \mu_0 \leq \half |\rho| (d-1)^2 \,, \hspace{1cm} \kappa_0^2 > \frac{(d-1)^2}{4}\Bigl(\mu_0 - \frac{(d-1)^2}{4} |\rho| \Bigr)\,,\hspace{1cm} IB.
\eeq
\noindent {\it Type II solutions for AdS:}
\beq
\label{man-12032017-03} && \mu_0 =  2 |\rho| (\lambda_0 + \frac{d-1}{2})^2\,, \hspace{1cm} \kappa_0^2 >   |\rho| (\lambda_0 + \frac{d-1}{2})^4\,, \qquad \lambda_0 = 1,2,\ldots, \infty\,.\qquad
\eeq
\noindent {\it Type III solutions for AdS:}
\vspace{-0.2cm}
\beq
\label{man-12032017-04} && \mu_0 = 2|\rho| (\lambda +\frac{d-1}{2})^2\,,
\\
\label{man-12032017-05} && \kappa_0^2 >   |\rho| (\lambda +\frac{d-1}{2})^4   -|\rho| \epsilon^2 (\epsilon + 2\lambda_0 + d-1)^2  \hspace{1.6cm} \hbox{ for } \ 0 < \epsilon < \epsilon_r\,,
\\
\label{man-12032017-06} && \kappa_0^2 >   |\rho| (\lambda +\frac{d-1}{2})^4  -|\rho| (1-\epsilon)^2 (\epsilon + 2\lambda_0 + d)^2\,,\hspace{1.1cm} \hbox{ for } \ \epsilon_r < \epsilon < 1\,,\qquad
\\
\label{man-12032017-07} && \kappa_0^2 >   |\rho|(\lambda +\frac{d-1}{2})^4 - \frac{1}{4}|\rho|(2\lambda_0 + d)^2\,,  \hspace{2.8cm} \hbox{ for } \ \epsilon = \epsilon_r\,,
\\
\label{man-12032017-08} && \lambda = \lambda_0+\epsilon\,, \qquad 0 < \epsilon < 1\,, \qquad \lambda_0=0,1,\ldots,\infty \,,
\\
\label{man-12032017-09} && \epsilon_r \equiv  \half\bigl( \sqrt{ (2\lambda_0 + d)^2 + 1} - 2\lambda_0 - d + 1\bigr).
\eeq
}
Note that type II solutions given in \rf{man-12032017-03} are labelled by integer $\lambda_0$, while type III solutions given in \rf{man-12032017-03}-\rf{man-12032017-09} are labelled by $\epsilon$,  $0<\epsilon<1$, and integer $\lambda_0$ \rf{man-12032017-08}.

\noindent {\bf Statement 3}. {\it For AdS, solution to Eqs.\rf{man-19022017-20} is given by}
\beq
\label{man-19022017-01} && \kappa_0^2 = \bigl(s+ \frac{d-1}{2}\bigr)^2 m^2\,, \qquad \mu_0 \equiv  m^2 +  |\rho| \bigl( s+\frac{d-1}{2} \bigr)^2 \,,
\\
\label{man-19022017-02} &&  |\rho|  \bigl(s + \frac{d-3}{2}\bigr)^2 < m^2  <   |\rho|  \bigl(s + \frac{d+1}{2}\bigr)^2\,,
\eeq
where, for the reader convenience, we introduce conventional mass parameter $m$ \rf{man-19022017-01}.%
\footnote{ For a finite component field, the $m$ is defined from the Lagrangian, ${\rm i}\LL = e \psi_s(\Dline + m)\psi_s + \ldots$ where dots stand for terms involving $\psi_{s'}$, $s'<s$, and
for contributions which vanish while imposing the $\gamma$-tracelessness constraint.
}
Lagrangian \rf{man-11022017-04} with  $\kappa_0$ and $\mu_0$ as in \rf{man-19022017-01}, \rf{man-19022017-02} describes the reducible classically unitary system. Namely, let us decompose ket-vectors $\psik$ \rf{man-11022017-03} and $\xik$ \rf{man-11022017-15} as
\beq
\label{man-19022017-04} && \psik = |\psi^{0,s}\rangle  + |\psi^{s+1,\infty}\rangle \,,
\\
\label{man-19022017-05} && \xik =  |\xi^{0,s-1}\rangle  + |\xi^{s,\infty}\rangle \,,
\eeq
where ket-vectors $|\psi^{M,N}\rangle$, $|\xi^{M,N}\rangle$ are defined in \rf{man-19022017-24},\rf{man-19022017-26}. Then we can verify that Lagrangian \rf{man-11022017-04} and gauge transformation \rf{man-11022017-16} with  $\kappa_0$ and $\mu_0$ as in \rf{man-19022017-01} are factorized as
\beq
\label{man-19022017-06} && \hspace{-1.5cm}\LL = \LL^{0,s} + \LL^{s+1,\infty}\,, \qquad \irm \LL^{0,s} \equiv e \langle \psi^{0,s}|E|\psi^{0,s}\rangle\,,  \qquad \irm \LL^{s+1,\infty} \equiv
e\langle \psi^{s+1,\infty}|E|\psi^{s+1,\infty}\rangle\,,\qquad
\\
\label{man-19022017-07} && \hspace{2.9cm} \delta |\psi^{0,s}\rangle  = G|\xi^{0,s-1}\rangle\,, \hspace{1.1cm} \delta |\psi^{s+1,\infty}\rangle  = G |\xi^{s,\infty}\rangle\,,
\eeq
where $E$, $G$ are given in \rf{man-11022017-05}, \rf{man-11022017-16}. In other words, $\LL^{0,s}$ and $\LL^{s+1,\infty}$ \rf{man-19022017-06} are invariant under transformations governed by gauge transformation parameters $|\xi^{0,s-1}\rangle$ and $|\xi^{s,\infty}\rangle$ respectively.

The three Statements above-discussed are proved by noticing that $F_\upsilon(n)$ \rf{man-19022017-14} has at most two roots. Thus we have to analyse the following three cases: 1) $F_\upsilon(n)$ has no roots; 2) $F_\upsilon(n)$ has one root; 3) $F_\upsilon(n)$ has two roots; We analyse these cases in turn.

\noindent \ibf) {\bf Solution without roots of $F_\upsilon(n)$}.  Using \rf{man-19022017-14} , we note that equations \rf{man-19022017-18} can alternatively be represented as
\beq
\label{man-15022017-04} && \kappa_0^2 > \max_{n=0,1,\ldots,\infty}  \Bigl(  \mu_0 \bigl( n + \frac{d-1}{2} \bigr)^2  +  \rho \bigl( n + \frac{d-1}{2} \bigr)^4 \Bigr)\,.
\eeq
We now see that, for dS space ($\rho > 0$), equation \rf{man-15022017-04} has no solution. For AdS space ($\rho <0$), we find that, depending on values of $\mu_0$, equation \rf{man-15022017-04} leads to Type IA, IB, II and III solutions.

\noindent \iibf) {\bf Solution with one root of $F_\upsilon(n)$}. Using the notation $n_r = s$ for one root of $F_\upsilon(n)$, we see that Eqs.\rf{man-19022017-20} amount to
\be  \label{man-20022017-07}
F_\upsilon(s) = 0\,, \qquad  F_\upsilon(n) > 0 \quad \hbox{ for } \quad n =0,1,\ldots, s-1, s+1,s+2,\ldots, \infty\,.
\ee
First we note that the equation $F_\upsilon(s) = 0$ leads to the following restrictions on $\mu_0$ and $\kappa_0$
\be \label{man-20022017-08}
\kappa_0^2 = \bigl(s+ \frac{d-1}{2}\bigr)^2 m^2\,, \qquad   \mu_0 \equiv  m^2 -  \rho \bigl( s+\frac{d-1}{2} \bigr)^2 \,.
\ee
Inserting $\mu_0$, $\kappa_0$ \rf{man-20022017-08} into \rf{man-11022017-10}, we cast the $F_\upsilon$ into the following form
\be \label{man-09022017-46}
F_\upsilon =   (s-N_\upsilon)(s+N_\upsilon +d-1) \Bigl( m^2 + \rho \bigl( N_\upsilon + \frac{d-1}{2} \bigr)^2 \Bigr) \,.
\ee
Lagrangian \rf{man-11022017-04} with $F_\upsilon$ given in \rf{man-09022017-46} describes reducible system of continuous spin field. This is to say that if we decompose $\psik$ and $\xik$ as in \rf{man-19022017-04}, \rf{man-19022017-05} then we can check that Lagrangian \rf{man-11022017-04} and gauge transformation \rf{man-11022017-16} are factorized as in \rf{man-19022017-06}, \rf{man-19022017-07}.

Second, using $\mu_0$, $\kappa_0$ \rf{man-20022017-08} and considering inequalities $F_\upsilon(n) > 0$ in \rf{man-20022017-07}, we find the following restrictions on $m^2$:
\beq
\label{man-20022017-09-b1} &&  |\rho|  (s + \frac{d-3}{2})^2 < m^2  <    |\rho|  (s + \frac{d+1}{2})^2 \hspace{2.9cm} \hbox{ for AdS};
\\
\label{man-20022017-09-b2} &&  - |\rho| \frac{(d-1)^2}{4}  < m^2 < - |\rho| \infty \,,  \hspace{4.5cm} \hbox{ for dS}.
\eeq
We note that for the derivation of inequalities in \rf{man-20022017-09-b1},\rf{man-20022017-09-b2} it is convenient to use representation for $F_\upsilon$ given in \rf{man-09022017-46}.
We note also that the left inequalities in \rf{man-20022017-09-b1},\rf{man-20022017-09-b2} are obtained by requiring $F_\upsilon(n)>0$ for $n=0,1,\ldots,s-1$, while the right inequalities in \rf{man-20022017-09-b1},\rf{man-20022017-09-b2} are obtained by requiring $F_\upsilon(n)>0$ for $n=s+1,\ldots, \infty$. It is easy to see that, for dS, inequalities \rf{man-20022017-09-b2} are inconsistent,%
\footnote{ The left inequality in \rf{man-20022017-09-b2} tells us that $m^2$ should be bounded from below, while from the right inequality in \rf{man-20022017-09-b2} we learn that $m^2 = - \infty$. Note that the right inequality in  \rf{man-20022017-09-b2} is obtained by requiring $F_\upsilon(n)>0$ for $n=\infty$. }
i.e., for dS, Eqs.\rf{man-19022017-20} with one root $n_r=s$ do not have solutions.
For AdS, we see that \rf{man-20022017-08}, \rf{man-20022017-09-b1} lead to \rf{man-19022017-01},\rf{man-19022017-02}.

\noindent \iiibf) {\bf Solution with two roots of $F_\upsilon(n)$}.  Denoting two roots of Eqs.\rf{man-19022017-20} as $s,S$,
\be  \label{man-20022017-14}
F_\upsilon(s) = 0 \,, \qquad F_\upsilon(S) = 0\,, \qquad s \leq S\,,
\ee
it is easy to see Eqs.\rf{man-20022017-14} amount to the following relations
\beq
\label{man-20022017-15} && \kappa_0^2 = -\rho \bigl( s+\frac{d-1}{2} \bigr)^2 \bigl( S+\frac{d-1}{2} \bigr)^2\,,
\\
\label{man-20022017-16} && \mu_0  = - \rho \bigl( s+\frac{d-1}{2} \bigr)^2 - \rho \bigl( S + \frac{d-1}{2} \bigr)^2\,.
\eeq
Inserting $\mu_0$, $\kappa_0$ \rf{man-20022017-15}, \rf{man-20022017-16} into \rf{man-11022017-10}, we find the following representation for $F_\upsilon$:
\be \label{man-09022017-50}
F_\upsilon =   -\rho (s-N_\upsilon)(s+N_\upsilon +d-1)(S-N_\upsilon)(S+N_\upsilon +d-1)  \,.
\ee
Lagrangian \rf{man-11022017-04} with $F_\upsilon$ as in \rf{man-09022017-50} describes reducible system of continuous spin field. Namely let us decompose ket-vectors $\psik$ \rf{man-11022017-03} and $\xik$ \rf{man-11022017-15} as
\beq
\label{man-19022017-04-a1} && \psik = |\psi^{0,s}\rangle  + |\psi^{s+1,S}\rangle + |\psi^{S+1,\infty}\rangle \,,
\\
\label{man-19022017-05-a1} && \xik= |\xi^{0,s-1}\rangle  + |\xi^{s,S-1}\rangle + |\xi^{S,\infty}\rangle\,,
\eeq
where the ket-vectors $|\psi^{M,N}\rangle$, $|\xi^{M,N}\rangle$ are defined in \rf{man-19022017-24},\rf{man-19022017-26}. Then we can verify that Lagrangian \rf{man-11022017-04} and gauge transformation \rf{man-11022017-16} with  $\kappa_0$ and $\mu_0$ as in \rf{man-20022017-15},\rf{man-20022017-16}  are factorized,
\beq
\label{man-12032017-10} && \LL = \LL^{0,s} + \LL^{s+1,S} + \LL^{S+1,\infty}\,, \qquad \irm\LL^{M,N} \equiv  e   \langle \psi^{M,N}|E|\psi^{M,N}\rangle\,,
\\
\label{man-12032017-11} &&  \delta |\psi^{0,s}\rangle  = G|\xi^{0,s-1}\rangle\,, \qquad \delta |\psi^{s+1,S}\rangle  = G|\xi^{s,S-1}\rangle\,,  \qquad \delta |\psi^{S+1,\infty}\rangle  = G |\xi^{S,\infty}\rangle\,,\qquad
\eeq
where $E$, $G$ are given in \rf{man-11022017-05}, \rf{man-11022017-16}. In \rf{man-12032017-10}, \rf{man-12032017-11}, we assume that if $s=S$, then  $|\psi^{s+1,s}\rangle\equiv 0$. Note that classically unitary system with $s=S$ is described by \rf{man-19022017-01},\rf{man-19022017-02} when $m^2 = |\rho|(s+\frac{d-1}{2})^2$.  From \rf{man-09022017-50}, we learn that the ket-vector $|\psi^{0,s}\rangle$ describes classically unitary (non-unitary) fermionic massive spin-$(s+\half)$ field in AdS (dS), the ket-vector $|\psi^{s+1,S}\rangle$ describes classically non-unitary fermionic partial-massless spin-$(S+\half)$ field in (A)dS, while the ket-vector $|\psi^{S+1,\infty}\rangle$ describes classically unitary (non-unitary) fermionic infinite-component spin-$(S+\frac{3}{2})$ field in AdS (dS). Mass parameters of the fermionic fields $|\psi^{0,s}\rangle$ and $|\psi^{s+1,S}\rangle$ are given by
\beq
\label{man-12032017-14} && m^2 = - \rho \bigl(S+ \frac{d-1}{2}\bigr)^2\,, \hspace{3cm} \hbox{ for } \ \ |\psi^{0,s}\rangle\,,
\\
\label{man-12032017-15} && m^2 = - \rho \bigl(s+ \frac{d-1}{2}\bigr)^2\,, \hspace{3.1cm} \hbox{ for } \ \ |\psi^{s+1,S}\rangle\,.
\eeq
Mass parameter \rf{man-12032017-15} can be represented in the form
\be \label{man-12032017-15-x1}
m^2 = - \rho \bigl(S+ \frac{d-3}{2} -k \bigr)^2\,, \qquad k \equiv S-s-1\,, \hspace{1cm} \hbox{ for } \ \ |\psi^{s+1,S}\rangle\,,
\ee
which tells that us that $|\psi^{s+1,S}\rangle$ describes a spin-$(S+\half)$ and depth-$k$ partial-massless fermionic field. For $S=s+1$ ($k=0$), this field turns out to be spin-$(S+\half)$ massless fermionic field.%
\footnote{ Using notation $k_{_{\rm CU}}$ for the commonly used depth of partial-massless field we note that our $k$ in \rf{man-12032017-15-x1} is related to $k_{_{\rm CU}}$ as $k= k_{_{\rm CU}}-1$.}

\newsection{\large  Partition function and modified de Donder gauge condition}\label{secti-03}

In this section, we are going to demonstrate that a partition function of the fermionic continuous spin field is equal to one. Namely, using gauge invariant Lagrangian \rf{man-11022017-04}, we find the following expression for the partition function $Z$ of the fermionic continuous spin field
\beq
\label{man-20022017-01} && Z^{-1} = \prod_{n=0}^\infty Z_n^{-1}\,, \qquad Z_n^{-1} = \frac{\DD_{n-1}(M_{n-1}^2)\DD_{n-1}(M_{n-1}^2)}{\DD_{n}(M_n^2)\DD_{n-2}(M_{n-2}^2)}\,,
\\
\label{man-20022017-03} && \DD_{n}(M^2) = \sqrt{\det{}_n(- \Boxline + M^2)}\,, \qquad \Boxline \equiv \Dline \Dline\,,
\\
\label{man-20022017-05} && M_n^2 \equiv \mu_0 + \rho \bigl(n+\frac{d-1}{2}\bigr)^2\,,
\eeq
where, in \rf{man-20022017-03}, a determinant is computed on a space of the Lorentz $so(d,1)$ algebra spin-$(n+\half)$ Dirac field subject to $\gamma$-tracelessness constraint. Note also that in \rf{man-20022017-01} we assume the convention $\DD_{-2}(M^2)=1$, $\DD_{-1}(M^2)=1$. From \rf{man-20022017-01}, we see immediately that the partition function of the fermionic continuous spin field is equal to one, $Z=1$.

Alternatively, $Z$ \rf{man-20022017-01} can be cast into more convenient form by using general formula for the $\DD_n(M^2)$ with arbitrary $M^2$ \rf{man-20022017-03},
\be \label{man-20022017-19}
\DD_n(M^2) = \DD_n^\perp(M^2) \DD_{n-1}(M^2 -  \rho(2n+d-2))\,,
\ee
where $\DD_n^\perp(M^2)$ appearing in \rf{man-20022017-19} takes the same form as in \rf{man-20022017-03}, with assumption that the determinant is computed on space of the gamma-traceless and divergence-free spin-$(n+\half)$ Dirac field of the Lorentz $so(d,1)$ algebra. For $M_n^2$ as in \rf{man-20022017-05}, relation \rf{man-20022017-19} takes the form
\be \label{man-20022017-20}
\DD_n(M_n^2) = \DD_n^\perp(M_n^2) \DD_{n-1}(M_{n-1}^2)\,.
\ee
Using \rf{man-20022017-20}, we see that $Z_n^{-1}$ \rf{man-20022017-01} can be represented as
\be  \label{man-15032017-01}
Z_n^{-1} =  \frac{\DD_{n-1}^\perp(M_{n-1})}{\DD_n^\perp(M_n)}\,.
\ee
Using \rf{man-15032017-01}, we find then, for fermionic continuous spin field in (A)dS and flat spaces, the same cancellation mechanism as for higher-spin fields in flat space \cite{Beccaria:2015vaa}.
Namely, from $Z$ \rf{man-20022017-01} and $Z_n$ \rf{man-15032017-01}, we see the cancellation of determinant of the physical spin-$(n+\half)$ field and ghost determinant of spin-$(n+\frac{3}{2})$ field. Note that $Z$ is equal to 1 without the use of special regularization procedure required for a computation of partition functions of higher-spin (A)dS gauge fields (see, e.g., Ref.\cite{Beccaria:2015vaa}).

Representation for the partition function given in \rf{man-20022017-01} can be obtained from Lagrangian \rf{man-11022017-04} by using the well-known technique discussed in the earlier literature (see, e.g., Refs.\cite{Creutzig:2011fe}-\cite{Campoleoni:2015qrh}). Here, instead of a repetition of the well-known technicalities, we prefer to demonstrate how the expression for partition function of continuous spin fermionic field \rf{man-20022017-01} leads to partition functions of massless, partial-massless and massive fermionic fields. Also we present our modified de Donder gauge for fermionic fields which allows us to obtain in a straightforward way the mass terms $M_n^2$ \rf{man-20022017-05} entering the determinants in \rf{man-20022017-01},\rf{man-20022017-03}.

\noindent {\bf Partial-massless and massless fields}. As we said above, if $\kappa_0$ and $\mu_0$ take values given in \rf{man-20022017-15}, \rf{man-20022017-16}, then field $|\psi^{s+1,S}\rangle$ appearing in \rf{man-19022017-04-a1} describes spin-$(S+\half)$, depth-$(S-s-1)$ partial-massless field. Partition function for such field is obtained from \rf{man-20022017-01} by considering contribution of $Z_n$ with $n=s+1,s+2,\ldots, S$. Doing so, we get
\be \label{man-20022017-17}
Z_{\rm par-massl}^{-1} = \prod_{n=s+1}^S Z_n^{-1}\,, \qquad Z_{\rm par-massl}^{-1} = \frac{ \DD_s (M_s^2) \DD_{\Ssm-1}(M_{\Ssm-1}^2) }{ \DD_{s-1}(M_{s-1}^2) \DD_\Ssm(M_\Ssm^2) }\,,\qquad
\ee
where $Z_n$ entering \rf{man-20022017-17} takes the same form as in \rf{man-20022017-01}, while $M_n^2$ is obtained by inserting $\mu_0$ \rf{man-20022017-16} into \rf{man-20022017-05}.
\be \label{man-20022017-18}
M_n^2   =  \rho \bigl( n + \frac{d-1}{2} \bigr)^2- \rho \bigl( s+\frac{d-1}{2} \bigr)^2 - \rho \bigl( S + \frac{d-1}{2} \bigr)^2\,.
\ee
Using \rf{man-20022017-19} and $M_n^2$ \rf{man-20022017-18}, we verify relation \rf{man-20022017-20}. Using \rf{man-20022017-20}, we see that $Z$ \rf{man-20022017-17} takes the well-known form of partition function of spin-$(S+\half)$ and depth-$(S-s-1)$ partial-massless field
\beq
\label{man-20022017-22} && \hspace{-0.5cm} Z_{\rm par-massl}^{-1} = \frac{ \DD_s^\perp (M_s^2) }{ \DD_\Ssm^\perp(M_\Ssm^2) }\,, \qquad M_s^2   =  - \rho \bigl( S + \frac{d-1}{2} \bigr)^2\,, \qquad M_S^2   =  - \rho \bigl( s+\frac{d-1}{2} \bigr)^2\,, \qquad
\eeq
where $M_s^2$, $M_S^2$ given in \rf{man-20022017-22} are obtained from \rf{man-20022017-18}. For $S=s+1$, a ket-vector $|\psi^{S,S}\rangle$ \rf{man-19022017-04-a1} describes massless field and $Z$ \rf{man-20022017-22} becomes the  partition function of spin-$(S+\half)$ massless field
\be  \label{man-20022017-23}
Z_{\rm massl}^{-1} = \frac{\DD_{\Ssm-1}^\perp(M_{\Ssm-1}^2)}{\DD_\Ssm^\perp(M_{\Ssm}^2)} \,, \qquad M_{\Ssm-1}^2 = - \rho (S+\frac{d-1}{2})^2 \qquad M_\Ssm^2 = - \rho (S+\frac{d-3}{2})^2\,.\qquad
\ee

\noindent {\bf Massive field}. As we noted, if $\kappa_0$ and $\mu_0$ take values given in \rf{man-20022017-08}, then field $|\psi^{0,s}\rangle$ appearing in \rf{man-19022017-04} describes spin-$(s+\half)$ and mass-$m$ massive field. Partition function for such field is obtained from \rf{man-20022017-01} by considering the contributions of $Z_n$ with $n=0,1,\ldots, s$. Doing so, we get
\be  \label{man-20022017-24}
Z_{\rm massv}^{-1} = \prod_{n=0}^s Z_n^{-1}\,, \qquad Z_{\rm massv}^{-1} = \frac{\DD_{s-1}(M_{s-1}^2) }{ \DD_s(M_s^2) }\,,\qquad
\ee
where $Z_n$ entering \rf{man-20022017-24} takes the same form as in \rf{man-20022017-01}, while $M_n^2$ is obtained by inserting $\mu_0$ \rf{man-20022017-08} into \rf{man-20022017-05},
\be \label{man-20022017-25}
M_n^2   =  m^2 - \rho ( s-n)( s+ n + d-1)\,.
\ee
Using \rf{man-20022017-19} and $M_n^2$ \rf{man-20022017-25}, we verify relation \rf{man-20022017-20}.
Using then \rf{man-20022017-20} for $n=s$, we find that $Z$ \rf{man-20022017-24} takes the well-known form of partition function for spin-$(s+\half)$ and mass-$m$ massive fermionic field
\be \label{man-20022017-26}
Z_{\rm massv} = \DD_s^\perp(M_s^2) \,, \qquad M_s^2 = m^2\,.
\ee

\noindent {\bf Modified de Donder gauge}. Now we outline the derivation of the mass terms $M_n^2$ \rf{man-20022017-05} entering the determinants in \rf{man-20022017-01},\rf{man-20022017-03}. To this end we propose to use a gauge condition which we refer to as modified de Donder gauge. A modified de Donder gauge is defined as follows
\beq
\label{man-13032017-01} && \hspace{-2cm} \Cb_\mod\psik  =  0 \,,
\\
\label{man-13032017-02} \Cb_\mod  & \equiv &  \Cb_\st - \frac{1}{2N_\alpha+d-1}(\gaalb + \half \gaal \bar\alpha^2) e_1^\Gammasm + \half \bar\alpha^2 e_1 - \frac{2N_\alpha+d-3}{2N_\alpha+d-1}\Pi_\bos^\smponetwo \eb_1\,,\qquad\qquad
\\
\label{man-13032017-03} &&  \Cb_\st \equiv \alphab D - \half \alpha D \alphab^2\,, \qquad \Pi_\bos^\smponetwo  \equiv  1 - \alpha^2 \frac{1}{2(2N_\alpha + d + 1)}\alphab^2\,,
\eeq
where the operators $e_1^\Gammasm$, $e_1$, $\eb_1$ are given in \rf{man-11022017-08},\rf{man-11022017-09}.
By analogy with Lagrangian \rf{man-11022017-04} and gauge transformation \rf{man-11022017-16}, the representation for modified de Donder gauge given in \rf{man-13032017-01}-\rf{man-13032017-03} is universal and valid for arbitrary theory of fermionic totally symmetric (A)dS fields (see our discussion below Eq.\rf{man-11022017-16}).

Using \rf{man-13032017-01}, we verify that first-order equations of motion obtained from Lagrangian \rf{man-11022017-04} lead to the following second-order equations for $\psik$:%
\footnote{ For the bosonic case, we expect that the results in Ref.\cite{Metsaev:1999ui} and Ref.\cite{Ponomarev:2010st} provide  descriptions of bosonic continuous spin field at level of light-cone gauge equations of motion and unfolding equations of motion respectively. Interesting recent discussion of unfolded approach may be found in Ref.\cite{Basile:2017mqc}.}
\be \label{man-13032017-04}
(\Boxline  - M^2  + \rho \alpha^2 \alphab^2)\psik = 0\,, \qquad M^2 \equiv \mu_0 + \rho \bigl(N_\upsilon + \frac{d-1}{2} \bigr)^2\,,
\ee
where $\Boxline$ is given in \rf{man-20022017-03}. From \rf{man-13032017-04}, we see that, as we have promised, for flat space ($\rho = 0$), the parameter $\mu_0$ can be interpreted as square of the conventional mass parameter $m$, $\mu_0 = m^2$.

Gauge-fixed equations of motion \rf{man-13032017-04} are invariant under on-shell leftover gauge
transformation that are obtained from \rf{man-11022017-16} by the substitution $\xik\rightarrow \Xik$, where the $\Xik$  satisfies the following equations of motion:
\be \label{man-13032017-05}
(\Boxline - M_\Xi^2) \Xik = 0 \,, \qquad M_\Xi^2 \equiv \mu_0 + \rho \bigl(N_\upsilon + \frac{d-3}{2} \bigr)^2\,.
\ee
We note that ket-vector $\Xik$ is obtained from \rf{man-11022017-15} by the substitution $\xi^{a_1\ldots a_n}\rightarrow \Xi^{a_1\ldots a_n}$.

To cast equations \rf{man-13032017-04} into decoupled form, we decompose triple $\gamma$-traceless ket-vector $\psik$, $(\gaalb)^3\psik=0$, into three $\gamma$-traceless ket-vectors $|\psi_\Ism\rangle$, $|\psi_\Gammasm\rangle$, and $|\psi_\IIsm \rangle$,
\beq
\label{man-13032017-06} &&  \hspace{-1cm} \psik = |\psi_\Ism\rangle  + \gaal |\psi_\Gammasm\rangle  + \alpha^2 |\psi_\IIsm \rangle \,,\qquad \gaalb  |\psi_\tau\rangle = 0 \,, \quad \tau=    I , \Gamma,  II \,,
\\
\label{man-13032017-07} && \hspace{-1cm} |\psi_\tau\rangle = \sum_{n=0}^\infty \frac{\upsilon^{n+\lambda_\tau}}{n!\sqrt{(n+\lambda_\tau)!}} \alpha^{a_1} \ldots \alpha^{a_n} \psi_\tau^{a_1\ldots a_n} |0\rangle\,, \qquad
\lambda_\Ism \equiv 0 \,, \quad \lambda_\Gammasm \equiv 1\,, \quad \lambda_\IIsm \equiv 2\,.\qquad
\eeq
Inserting $\psik$ \rf{man-13032017-06} into \rf{man-13032017-04}, we obtain decoupled equations for $|\psi_\tau\rangle$, $\tau=I,\Gamma,II$:
\beq
\label{man-13032017-08} && (\Boxline - M_\tau^2)|\psi_\tau\rangle =0\,, \qquad  M_\tau^2 = \mu_0 + \rho \bigl( N_\upsilon + \frac{d-1}{2} - \lambda_\tau \bigr)^2\,.
\eeq
In terms of $\psi_\tau^{a_1\ldots a_n}$ and $\Xi^{a_1\ldots a_n}$, equations \rf{man-13032017-08} and \rf{man-13032017-05} take the following respective form
\be \label{man-13032017-09}
(\Boxline - M_n^2)\psi_\tau^{a_1\ldots a_n} = 0 \,, \quad \tau=I,\Gamma,II, \qquad (\Boxline - M_n^2)\Xi^{a_1\ldots a_n} = 0 \,,
\ee
$n=0,1,\ldots \infty$, where the mass terms $M_n^2$ take the form given in \rf{man-20022017-05}. Thus we see that the mass terms entering determinants \rf{man-20022017-01},\rf{man-20022017-03} are indeed governed by $M_n^2$  \rf{man-20022017-05}.

To conclude, we developed Lagrangian gauge invariant formulation for fermionic continuous
spin field in (A)dS. We used our formulation to demonstrate that a partition function of the
fermionic continuous spin field is equal to 1.
Lagrangian descriptions of bosonic and fermionic continuous spin (A)dS fields obtained in Ref.\cite{Metsaev:2016lhs} and in this paper provide possibility to discuss supersymmetric theories of continuous spin (A)dS fields. Recent interesting discussion of Lagrangian formulation for supermultiplets in AdS and flat spaces may be found in Refs.\cite{Buchbinder:2015mta,Kuzenko:2016qwo}. Needless to say that a problem of interacting continuous spin fields also deserves to be studied. Various light-cone gauge methods discussed in Refs.\cite{Bengtsson:1983pd}-\cite{Bengtsson:2016hss} might be helpful for this purpose. Use of Lorentz covariant methods (see, e.g., Refs \cite{Vasilev:2011xf}-\cite{Manvelyan:2010je}) for study of interacting continuous spin fields could also be of some interest. Our modified de Donder gauge leads to simple equations of motion for fermionic fields. We think therefore that such gauge might be useful for studying various aspects of (A)dS field dynamics along the lines in Refs.\cite{Ponomarev:2016jqk,Francia:2007qt}.

\medskip
{\bf Acknowledgments}. This work was supported by the RFBR Grant No. 17-02-00317.

\setcounter{section}{0} \setcounter{subsection}{0}
\appendix{ \large  Notation }

Vector indices of the Lorentz $so(d,1)$ algebra $a,b,c$ run over values $0,1,\ldots
,d$. Flat metric tensor $\eta^{ab}$ is mostly positive. For simplicity, we drop the flat metric $\eta^{ab}$ in scalar products.
The vacuum $|0\rangle$, the creation operators $\alpha^a$, $\upsilon$, and the
annihilation operators $\alphab^a$, $\upsilonb$ are defined by the relations
\be \label{man-14032017-01}
[\bar{\alpha}^a,\alpha^b]=\eta^{ab}\,,  \quad [\upsilonb,\upsilon]=1\,, \quad
\bar\alpha^a |0\rangle = 0\,,\quad  \upsilonb |0\rangle = 0\,, \qquad \alpha^{a\dagger} = \alphab^a, \qquad \upsilon^{\dagger} = \upsilonb.
\ee
Throughout this paper, we refer to these operators as oscillators. For the Dirac $\gamma$-matrices we use the conventions $\{\gamma^a,\gamma^b\} = 2\eta^{ab}$, $\gamma^{a\dagger} = \gamma^0 \gamma^a \gamma^0$.
On a space of ket-vector $\psik$ \rf{man-11022017-03}, covariant derivative $D^a$  is defined as $D^a = \eta^{ab}D_b$,
\be \label{man-14032017-02}
D_a \equiv e_a^\mun D_\mun\,,  \qquad D_\mun \equiv
\partial_\mun +\frac{1}{2}\omega_\mun^{ab} M^{ab}\,, \qquad   M^{ab} \equiv \alpha^a
\alphab^b + \frac{1}{4}\gamma^a\gamma^b - (a\leftrightarrow b)\,,
\ee
$\partial_\mun = \partial/\partial x^\mun$. In \rf{man-14032017-02}, the base manifold index $\mun$ run over values  = $0,1,\ldots, d$, the $D_\mun$ denotes Lorentz covariant derivative, the $e_a^\mun$ is inverse vielbein of $(A)dS_{d+1}$ space-time, the $M^{ab}$ stands for a spin
operator of the Lorentz algebra $so(d,1)$, and the $\omega_\mun^{ab}$ denotes the Lorentz connection of $(A)dS_{d+1}$ space-time. Tensor-spinor field carrying the base manifold indices, $\psi^{\mun_1\ldots \mun_n}$, and tensor-spinor field carrying the flat indices, $\psi^{a_1\ldots
a_n}$, are related as $\psi^{a_1\ldots a_n} \equiv e_{\mun_1}^{a_1}\ldots
e_{\mun_n}^{a_n} \psi^{\mun_1\ldots \mun_n}$. Our conventions for scalar products are as follows
\beq
\label{man-14032017-03} && \alpha^2 \equiv \alpha^a\alpha^a\,, \hspace{1cm} \alphab^2 \equiv \alphab^a\alphab^a\,,  \hspace{1cm} N_\alpha \equiv \alpha^a \alphab^a\,,  \hspace{1cm} N_\upsilon \equiv \upsilon \upsilonb\,,
\\
\label{man-14032017-04} && \gaal\equiv \gamma^a\alpha^a\,, \qquad \gaalb\equiv \gamma^a\alphab^a\,,\qquad \alpha D \equiv \alpha^a D^a\,, \hspace{1cm} \alphab D \equiv \alphab^a D^a\,, \hspace{0.5cm} \Dline \equiv \gamma^a D^a\,.\qquad
\eeq

\end{document}